\documentstyle[12pt,axodraw]{article}

\newcommand{\be}{\begin{equation}}
\newcommand{\bea}{\begin{eqnarray}}
\newcommand{\ee}{\end{equation}}
\newcommand{\eea}{\end{eqnarray}}

\def \qsl {q \kern-.45em{/}}
\def \slp {p \kern-.45em{/}}
\def \ksl {k \kern-.45em{/}}
\def \esl {\mbox{$\epsilon$} \kern-.45em{/}}
\def\v{\beta}
\def\c{s_w}

\catcode`@=11
\newcount\@tempcntc
\def\@citex[#1]#2{\if@filesw\immediate\write\@auxout{\string\citation{#2}}\fi
  \@tempcnta\z@\@tempcntb\m@ne\def\@citea{}\@cite{\@for\@citeb:=#2\do
    {\@ifundefined
       {b@\@citeb}{\@citeo\@tempcntb\m@ne\@citea\def\@citea{,}{\bf
?}\@warning
       {Citation `\@citeb' on page \thepage \space undefined}}%
    {\setbox\z@\hbox{\global\@tempcntc0\csname
b@\@citeb\endcsname\relax}%
     \ifnum\@tempcntc=\z@ \@citeo\@tempcntb\m@ne
       \@citea\def\@citea{,}\hbox{\csname b@\@citeb\endcsname}%
     \else
      \advance\@tempcntb\@ne
      \ifnum\@tempcntb=\@tempcntc
      \else\advance\@tempcntb\m@ne\@citeo
      \@tempcnta\@tempcntc\@tempcntb\@tempcntc\fi\fi}}\@citeo}{#1}}
\def\@citeo{\ifnum\@tempcnta>\@tempcntb\else\@citea\def\@citea{,}%
  \ifnum\@tempcnta=\@tempcntb\the\@tempcnta\else
   {\advance\@tempcnta\@ne\ifnum\@tempcnta=\@tempcntb \else
\def\@citea{--}\fi
    \advance\@tempcnta\m@ne\the\@tempcnta\@citea\the\@tempcntb}\fi\fi}
\catcode`@=12
 
\begin{document}

\begin{flushright}
FTUV-2000-2-16\\
February 2000
\end{flushright}
 
\begin{center}
{\Large {\bf A method for determining anomalous gauge \\ 
boson couplings from $e^{+}e^{-}$ experiments}}
\\[2.4cm]
{\large Joannis Papavassiliou}$^a$~ and 
{\large Kostas Philippides}$^b$\\[0.4cm]
$^a${\em 
Departamento de F\'{\i}sica Te\'orica, Univ. Valencia \\
E-46100 Burjassot (Valencia), Spain}\\[0.3cm]

$^b${\em Department of Theoretical Physics, 
Aristotelian University, \\ 
54006 Thessaloniki, Greece}\\[0.3cm]

\end{center}
 
\vskip0.7cm     \centerline{\bf   ABSTRACT}  \noindent

\vskip0.7cm

We present a  model-independent method for determining anomalous gauge
boson couplings from ongoing and  future $e^{+}e^{-} \to W^{+} W^{-} $
experiments.  First we generalize   an already existing  method, which
relies on the      study  of four  observables   constructed   through
appropriate projections of the unpolarized differential cross-section.
In particular, we   retain  both linear and   quadratic  terms in  the
unknown  couplings,  and compute  contributions  to  these observables
originating from anomalous couplings which  do not separately conserve
the discrete  $C$, $P$,  and $T$ symmetries.   Second,  we combine the
above set  of  observables with three  additional  ones, which can  be
experimentally obtained from the   total cross-sections for  polarized
final state $W$  bosons.  The resulting  set of  seven observables may
provide useful information  for  constraining, and  in  some cases for
fully determining,  various of   the  possible anomalous   gauge boson
couplings.

\newpage

\setcounter{equation}{0}
\section{Introduction}

Anomalous gauge  boson couplings  \cite{GG,Hagi}
have attracted significant attention
in   recent  years,  and their    direct  study  through  the  process
$e^{+}e^{-}\rightarrow  W^{+}W^{-}$ has been one of
the  main objectives  of the CERN   Large  Electron Positron  collider
LEP2 \cite{BBL,Var,HISZ,effthe,WG1,WG2}. 
In addition, the trilinear gauge self-couplings have also been
probed through direct $W\gamma$ and   $WZ$ production at the  Tevatron
\cite{EW,Hadron}.  The study of such couplings is expected to continue 
at the  CERN  Large Hadron Collider (LHC),  as   well as the  Next Linear
Collider (NLC) \cite{lin}.

Recently a model-independent method has been proposed for extracting
values   or   bounds for  the  anomalous   gauge  boson couplings from
$e^{+}e^{-}\rightarrow  W^{+}W^{-}$
experiments   \cite{PP}.  The  basic  idea  is  to  study
projections of the  differential cross-section which arise when the
latter is convoluted with  a set  of appropriately  constructed
polynomials in $\cos\theta$, where $\theta$ is the center-of-mass
scattering  angle. This construction   leads to  a  set of four
novel observables, which are related to the anomalous couplings by
means  of simple  algebraic  equations. The experimental
determination of these  observables can  in turn be used in order
to impose bounds  simultaneously   on all anomalous    couplings,
without having to resort    to  model-dependent   relations among
them, or invoke any further simplifying assumptions.  This method
has also been generalized to the  case of hadron colliders \cite{PP2},
and its compatibility with the inclusion of structure function effects
has been established.  In what follows we will refer to this method as
the ``Projective Method'' (PM).

The PM as  presented in \cite{PP}  only includes  terms linear  in the
unknown anomalous   couplings (form-factors) which   are  individually
invariant under the  discrete $C$, $P$, and  $T$ symmetries.  However,
the inclusion of the quadratic terms as well as the  $C$, $P$, and $T$
non-invariant couplings  is   necessary for  a  complete  experimental
analysis.  In addition, the observables constructed by means of the PM
are  only four,   whereas  the unknown   form-factors,  even with  the
simplifications  mentioned above, are  six; therefore,  
one  is not able to
extract experimental information for all anomalous couplings, but only
for a few   of  them.  In addition,   the  fact that  the  observables
constructed  by  this method are  rather  strongly  correlated further
reduces the predictive power of the PM.

The purpose of this paper is two-fold: 

(i) The contribution of {\it all} anomalous couplings 
is computed, and terms linear and quadratic are retained.
This not only augments the PM, but as we will see later,
results in the additional advantage of reducing the correlation
among the four original PM observables.

(ii) The  aforementioned four observables  of the PM are combined with
three  additional  observables, which can be extracted, 
at least in principle,
from  measurements of
polarized total cross-sections. Specifically,
the observables correspond to the total cross-section for 
having two transverse, two longitudinal, and one transverse
and one longitudinal $W$ bosons in the final state.
These quantities have already been studied 
in the literature  \cite{Bil}, and are usually denoted by
$\sigma_{TT}$, $\sigma_{LL}$, and 
$\sigma_{TL}$, respectively. In fact,
it is expected that experimental values for
the aforementioned observables can 
be extracted from the available LEP2 experimental
data \cite{RR}.  

The inclusion of $\sigma_{TT}$, $\sigma_{LL}$, and 
$\sigma_{TL}$ to the original PM observables
gives rise to a set  of seven observables, thus
increasing the predictive power of the method.  For practical purposes
in this  paper  we present the  case   where the aforementioned  three
cross-sections are  calculated keeping linear  and quadratic  parts of
the $C$, $P$, and $T$ invariant couplings only.
To the best of our knowledge 
the explicit closed form of these cross-sections in terms of the
anomalous couplings is presented here for the first time.

The outline  of the paper  is as follows:  In section 2 we present the
complete expressions for  the  PM observables, keeping all   terms. In
section  3  we compute the closed    expressions for the  polarized
cross-sections,   keeping    quadratic correction  but,   assuming the
presence  of $C$, and $P$  invariant couplings only.   In section 4 we
focus  on the  system of  equations obtained when  the results  of the
previous two  sections are  combined;  at this  stage the   linear and
quadratic terms  of the $C$ and $P$  invariant couplings are retained,
giving rise to seven  equations for six  unknown couplings. We discuss
various issues,   and  carry  out    an elementary analysis     of the
correlations among  some of these observables.   Finally, in section 5
we present our conclusions.

\setcounter{equation}{0}
\section{The complete expressions for the $\sigma_i$ observables }

In this section we extend the analysis presented in \cite{PP} by 
including the linear {\it and} quadratic contributions 
of {\it all} 
possible anomalous couplings. 
We consider the process 
$e^-(k_1,\sigma_1)e^+(k_2,\sigma_2) \to
W^-(p_1,\lambda_1)W^+(p_2,\lambda_2)$, shown in Fig.~1
The electrons are assumed to be massless, $\sigma_i$ 
label the spins of
the initial electron and  positron, i.e. 
$\sigma_1=-\sigma_2=\sigma/2,~\sigma=\pm 1$, whereas the  
$\lambda_i$ 
label 
the 
the polarizations of the produced $W$ bosons, with
$\lambda_i=0,\pm 1$.

The relevant kinematical variables
in the center-of-mass frame are 
\bea
s  = & (k_1+k_2)^2  = &(p_1+p_2)^2 ~,\nonumber\\
t  = & (k_1-p_1)^2  = &(p_2-k_2)^2 = 
-\frac{s}{4}(1+\beta^2-2 \beta \cos\theta)=-\frac{s\beta}{2}(z-x)~,
\label{defst}
\eea
where 
\be
 \beta = \sqrt{1-\frac{4M_W^2}{s}}~,
\ee
is the velocity of the $W$ bosons, $x\equiv \cos\theta$,  where $\theta$ is 
the angle between the incoming electron and the outgoing $W^-$ in the
center of mass frame, and
\be
 z=\frac{1+\beta^2}{2\beta}~.
\ee

We now proceed to compute the unpolarized differential cross-section
$(d\sigma/dx)$
corresponding to this process, i.e. we average over the 
initial spins and sum over the final polarizations. We have 
\be
\frac{d\sigma}{dx} 
= \bigg(\frac{1}{2s} \bigg) \bigg (\frac{\beta}{16\pi} \bigg )
\frac{g^4}{4} 
\sum_{\sigma,\lambda_1,\lambda_2}
|{\cal M}^{\sigma}(\lambda_1,\lambda_2)|^2 ~.
\ee
The first fraction is the flux factor, the second is a phase space factor,
and the factor of $1/4$ is due to the averaging over the initial helicities. 
All conventions are identical to those of \cite{PP} except that we have now 
pulled out the overall coupling constant factor and have denoted the 
remaining sum of amplitudes by ${\cal M}$. 

The $VW^{+}W^{-}$ vertex $\Gamma_{\mu\alpha\beta}^{V}$~~($V=\gamma,~Z$) 
we use has the form
\be
\Gamma_{\mu\alpha\beta}^{V}=\Gamma_{\mu\alpha\beta}^{0}+
\delta\Gamma_{\mu\alpha\beta}^{V}~, 
\ee
where
\be
\Gamma_{\mu\alpha\beta}^{0}(q,-p_1,-p_2) = 
(p_2-p_1)_{\mu}g_{\alpha\beta} +
2(q_{\beta}g_{\mu\alpha}-q_{\alpha}g_{\beta\mu})
\ee
is the canonical Standard Model (SM) 
three-boson vertex at tree-level, assuming that
the two $W$-bosons are on-shell, and thus dropping terms proportional
to $p_{1\alpha}$ and $p_{2\beta}$. The term $\delta\Gamma_{\mu\alpha\beta}^{V}$
contains 
all possible deviations from the SM canonical form, 
 compatible with Lorentz invariance, i.e.
\bea
\delta\Gamma_{\mu\alpha\beta}^{V}(q,-p_1,-p_2)&=&
f_1^{V}(p_2-p_1)^{\mu}g^{\alpha\beta} -
\frac{f_2^{V}}{2 M_W^2}q^{\alpha}q^{\beta}(p_2-p_1)^{\mu}
\nonumber\\
&&+2f_3^{V}(q^{\beta}g^{\mu\alpha}-q^{\alpha}g^{\beta\mu})
\nonumber\\
&&+if_4^{V}(q^{\beta}g^{\mu\alpha}+q^{\alpha}g^{\beta\mu})
+if_5^V\epsilon^{\mu\alpha\beta\rho}(p_2-p_1)_{\rho}
\nonumber\\
&&+f_6^V\epsilon^{\mu\alpha\beta\rho}q_{\rho}
+\frac{f_7^V}{M_W^2}(p_2-p_1)^{\mu}\epsilon^{\alpha\beta\rho\sigma}
q_{\rho}(p_2-p_1)_{\sigma}~.
\label{anomalia}
\eea
The deviation 
form-factors $f^V_i$ are all zero in SM. 
In what follows they will also be referred
to as trilinear couplings or anomalous couplings. 
We assume all anomalous couplings to be real.

The calculation is straightforward but lengthy;
it is important to emphasize
that the inclusion of the additional terms in the vertex, namely those
that are not separately $C$ and $P$ invariant 
($f^V_4,f^V_5,f^V_6,f^V_7$) does {\it not } change the
functional dependence of the differential cross-section 
on the center-of-mass angle $\theta$, which was established
in \cite{PP}.
Thus, the expression for $(d\sigma_{an}/dx)$, 
the part 
of the differential cross-section which contains the
anomalous couplings, 
assumes again the form 
\be
(z-x)\frac{d\sigma_{an}}{dx}=\frac{g^4}{64\pi}\frac{\beta}{s}\ 
\sum_{i=1}^{4}\sigma_i(s) P_i(s,x) 
\label{an}
\ee
with  $P_i(s,x)$  the same polynomials
 in $x$ first obtained
in \cite{PP} namely :
\bea
P_1(x)&=&z-x~,\nonumber\\
P_2(x)&=&(z-x)(1-x^2)~,\nonumber\\
P_3(x)&=&1-x^2~,\nonumber\\
P_4(x)&=&1-\beta x~.
\eea
In arriving at this result the following algebraic identities 
\bea
x(z-x)&=&-\frac{1}{\beta}P_1+P_3+\frac{1}{2\eta\beta^2}P_4 \nonumber\\
x-\beta&=&-2P_1+\frac{1}{\beta}P_4
\eea
may be found useful.

Notice that the
explicit closed expressions of the coefficients 
$\sigma_i$ have changed with respect to those reported in \cite{PP},
since both the {\it linear} as well
as the {\it quadratic} dependence on {\it all} couplings has now
been included. Using the following uniform short-hand notation 
\be
c_{1,...,7}\equiv f^{\gamma}_{1,...,7}~,~~~~
c_{8,...,14}\equiv f^{Z}_{1,...,7}~,
\ee
we have that the $\sigma_i$, where $i=1,...4$, can be written as
\be
\sigma_i= \sum_{k}L^i_k\  c_k +
\sum_{k}\sum_{\ell\geq k} Q^i_{[k][\ell]}\ c_k c_{\ell}~.
\label{sLQ}
\ee

Defining the following abbreviations 
\be
\eta \equiv \frac{s}{4M_W^2},~~~u \equiv \frac{s}{s-M_Z^2},~~~
y \equiv \frac{u}{c_w^2},~~~r\equiv v^2+a^2~,
\ee
the explicit forms of the coefficients
$L^i_k$ and $Q^i_{[k][\ell]}$ of the linear and quadratic
 terms respectively in Eq.(\ref{sLQ})  are given below :

\bea
\begin{array}{ll}
          L^{1}_{3} =-8\c^2 [4\c^2 + v (4c_w^{2}-1)y ]~,   &
          L^1_{5}   = 4\c^2 [  a (4c_w^{2}-1)y -1 ]~, \\
          L^1_{10}  = -8u [4v\c^2 + r(4c_w^{2}-1) y]~,  &
          L^1_{12}  = -4(v+a)u + 8au[va(4c_w^{2}-1)y + 2\c^2 ]~.
\end{array}
\eea

\medskip

\bea 
\begin{array}{ll}
            Q^{1}_{[3][3]} ~~=  16 \c^4 \eta \v^2~,  &
            Q^{1}_{[3][10]} ~=  32v \c^2\eta\v^2u~, \\
            Q^{1}_{[3][12]} ~= -16a\c^2\eta\v^2 u~, &
            Q^1_{[4][4]} ~~= 4 \c^4 \eta \v^2~, \\
            Q^1_{[4][11]} ~= 8 v \c^2 \eta \v^2 u~, &
            Q^1_{[4][13]} ~= -8a\c^2\eta u~,\\   
            Q^1_{[5][5]}  ~~= 4 \c^4 \eta \v^4~, &
            Q^1_{[5][10]} ~= -16a\c^2\eta\v^2 u~, \\
            Q^1_{[5][12]} ~=  8 v \c^2 \eta \v^4 u~, &
            Q^1_{[6][6]}  ~~=  4 \c^4 \eta~, \\
            Q^1_{[6][11]} ~=  -8a\c^2\eta u~, &
            Q^1_{[6][13]} ~= 8 v \c^2 \eta u~, \\
            Q^1_{[10][10]} = 16\eta \v^2 r u^2~, &
            Q^1_{[10][12]} = -32va\eta\v^2 u^2~, \\
            Q^1_{[11][11]} = 4\eta \v^2 r u^2~, &
            Q^1_{[11][13]} = -16va\eta u^2~,\\
            Q^1_{[12][12]} = 4 \eta \v^4 r u^2~, &
            Q^1_{[13][13]} = 4\eta r u^2~. 
\end{array}
\eea 

\bea
\begin{array}{ll}
            L^2_{1} = \c^2\v^2 [2(3-2\eta)(vu+\c^2)-(1+2\eta)vy]~, &
            L^2_{2} = -2\beta^2\eta s_w^2 [2(1+\eta)(vu+ s_w^2) -\eta y v]~, \\
           L^2_{3}  = -4\beta^2\eta s_w^2 [2(vu+ s_w^2) -yv]~, &
           L^2_{8} = \beta^2 u 
           \left[2(3-2\eta)( ru+vs_w^2 )-(1+2\eta)y r\right] \\
            L^2_{9}  = -2\beta^2\eta u [2(1+\eta)(ru+ v s_w^2)-\eta y r]~, &
            L^2_{10} = - 4\beta^2\eta u[2(ru+v s_w^2)-yr]~,
\end{array}
\eea
 
\bea
\begin{array}{ll}
             Q^2_{[1][1]} = \c^4 \eta^2 \v^2 (3-2\v^2+3\v^4)~, &
             Q^2_{[1][2]}  = -4\c^4 \eta^3 \v^4(1+\v^2)~, \\
             Q^2_{[1][3]}  = -8 \c^4 \eta^2 \v^2(1+\v^2)~, &
             Q^2_{[1][8]}  = 2 v\c^2 \eta^2 \v^2(3-2\v^2+3\v^4)u~, \\
             Q^2_{[1][9]}  = -4 v\c^2 \eta^3 \v^4(1+\v^2)u~, &
             Q^2_{[1][10]} = -8 v\c^2\eta^2\v^2(1+\v^2)u~, \\
             Q^2_{[2][2]} = 4 \c^4 \eta^4 \v^6~,  &
             Q^2_{[2][3]} = 16\c^4 \eta^3 \v^4~, \\
             Q^2_{[2][8]}  =- 4 v\c^2 \eta^3 \v^4 (1+\v^2)u~,  &
             Q^2_{[2][9]} = 8 v\c^2 \eta^4 \v^6 u~, \\
             Q^2_{[2][10]} = 16 v\c^2 \eta^3 \v^4 u~,  &
             Q^2_{[3][3]} = 8 \c^4 \eta^2 \v^2 (1+\v^2)~, \\ 
             Q^2_{[3][8]}  = -8 v \c^2 \eta^2  \v^2 (1+\v^2) u~,  &
             Q^2_{[3][9]}  = 16 v\c^2 \eta^3\v^4 u~, \\
             Q^2_{[3][10]} = 16 v\c^2 \eta^2 \v^2 (1+\v^2)u~,  &
             Q^2_{[4][4]}   = -2 \c^4 \eta \v^2~,  \\
             Q^2_{[4][11]}  = -4 v \c^2 \eta \v^2 u~, &
             Q^2_{[5][5]}  = - 2 \c^4 \eta\v^4~,  \\
             Q^2_{[5][12]}  = -4 v \c^2 \eta\v^4 u~,  &
             Q^2_{[6][6]}   = -2 \c^4\eta\v^2~,\\ 
             Q^2_{[6][7]}  = 16 \c^4 \eta \v^2~,  &
             Q^2_{[6][13]} = -4 v \c^2 \eta \v^2 u~,  \\
             Q^2_{[6][14]} = 16 v \c^2\eta \v^2 u~,  &  \\ 
             Q^2_{[7][7]}  = 32 \c^4 \eta^2 \v^4~,  &
             Q^2_{[7][13]} = 16 v \c^2 \eta \v^2 u~,  \\
             Q^2_{[7][14]} = 64 v \c^2 \eta^2 \v^4 u~, &
             Q^2_{[8][8]}  = \eta^2 \v^2 r (3-2\v^2+3\v^4) u^2~,\\
             Q^2_{[8][9]} = -4 \eta^3 \v^4 r (1+\v^2) u^2~, &
             Q^2_{[8][10]} = -8 \eta^2 \v^2 r (1+\v^2) u^2~,   \\
             Q^2_{[9][9]} = 4\eta^4 \v^6 r u^2~, &
             Q^2_{[9][10]} = 16\eta^3 \v^4 r u^2~,  \\
             Q^2_{[10][10]} = 8\eta^2 \v^2 r (1+\v^2) u^2~, &
             Q^2_{[11][11]} = -2\eta \v^2 r u^2 ~, \\
             Q^2_{[12][12]} = -2\eta\v^4 r u^2~,  &
             Q^2_{[13][13]} = -2 \eta \v^2 r u^2 ~, \\
             Q^2_{[13][14]} = 16 \eta \v^2 r u^2 ~, &           
             Q^2_{[14][14]} = 32 \eta^2 \v^4 r u^2~.
\end{array}
\eea

\bea       
\begin{array}{ll}            
            L^3_{1}  = -\c^2 \v~,  &
            L^3_{2}  = \c^2 \eta\v~, \\ 
            L^3_{5}  = \c^2 \v [1 -4 a (4c_w^{2}-1) y]~, &
            L^3_{8}  = -(v+a)\v u~, \\
            L^3_{9}  = (v+a)\eta\v u~,  &
            L^3_{12} = \v u [(v+a) -8va (4c_w^{2}-1) y - 16a\c^2]~. 
\end{array}
\eea 
\medskip  

\bea
\begin{array}{ll}
                 Q^3_{[3][12]} = 16a\c^2\eta\v^3 u~, &
                 Q^3_{[4][13]} = 8a\c^2\eta\v u~, \\
                 Q^3_{[5][10]} = 16a\c^2\eta\v^3 u~, &
                 Q^3_{[6][11]}  = 8a\c^2\v \eta u~, \\
                 Q^3_{[10][12]} = 32va\eta\v^3 u^2~, &
                 Q^3_{[11][13]} = 16va\eta\v u^2~.
\end{array}         
\eea

\bea 
\begin{array}{ll}       
         L^4_{3}  = 4\c^2 \beta^{-1}~,   &
         L^4_{5}  = -4 a\c^2 (4c_w^{2}-1)(z-\v)y + 2\c^2\beta^{-1}~, \\ 
         L^4_{10} = 4(v+a)u\beta^{-1}~, &
     L^4_{12} = -8(z-\v)[va (4c_w^{2}-1)y + 2a\c^2 ]u + 2(v+a)u\beta^{-1}~. 
\end{array}
\eea

\medskip

\bea 
\begin{array}{ll} 
             Q^4_{[3][12]} ~= 8a\c^2\v u~, &
             Q^4_{[4][13]} ~= 4a\c^2\v^{-1}  u~, \\
             Q^4_{[5][10]} ~= 8a\c^2\v u~, &
             
             Q^4_{[6][11]} ~= 4a\c^2 \v^{-1} u~, \\
             Q^4_{[10][12]} = 16va\v u^2~, &
             Q^4_{[11][13]} = 8va\v^{-1} u^2~. 
\end{array}
\eea 
 
\medskip

As explained in \cite{PP} the four quantities $\sigma_i$ constitute 
a set of observables; their experimental values 
may be obtained through an
appropriate convolution of the 
experimentally measured
unpolarized differential cross-section $d\sigma^{(exp)}/dx$ 
with   
a set of four polynomials, $\widetilde{P}_i(x)$, which are
orthonormal 
to the $P_i(x)$, i.e. they satisfy 
\be
\int_{-1}^{1}\widetilde{P}_i(x,s)P_j(x,s) dx = \delta_{ij} ~.
\ee
Clearly the set $\widetilde{P}_i$ is not uniquely determined;
in \cite{PP} we have reported the set with the lowest
possible power in $x$, namely : 
\bea
\widetilde{P}_1(x,s) &=& \frac{\eta}{2} (3 \beta +15x - 15 \beta x^2
 - 35 x^3) ~, \nonumber\\
\widetilde{P}_2(x,s)& =& \frac{35}{8} (-3x+5x^3) ~,\nonumber\\
\widetilde{P}_3(x,s)& =& \frac{5}{8}(3 +21z x  -9x^2-35z x^3 ) ~, \nonumber\\
\widetilde{P}_4(x,s)& =&
\frac{\eta}{2}(-3  -15 z x  + 15 x^2 + 35z x^3 )~.
\eea
In particular, the $\sigma_i^{(exp)}$ are given by
\be
\sigma_i^{(exp)} 
= \left[\frac{64\pi s}{g^4\beta}\right] 
 \int_{-1}^{1} dx (z-x)  
\Bigg (\frac{d\sigma^{(exp)}}{dx}-
\frac{d\sigma^{(0)}}{dx}\Bigg)\widetilde{P}_i(x,s) ~,
\label{SIG}
\ee
where $d\sigma^{(0)}/dx$ is the SM expression for the differential
cross-section in the absence of anomalous couplings \cite{Buras}.
Given the experimental measurement of the differential cross-section
${d\sigma^{(exp)}}/{dx}$ for {\it on shell} $W$s the four numbers
$\sigma_i$ can be extracted together with their related errors.
Subsequently Eq.(\ref{sLQ}) can be viewed as a system of four quadratic
equations with fourteen unknowns which although cannot be solved,
it appears feasible that it could be fitted for all couplings simultaneously
in a model independent way. In fact, using $U(1)$ electromagnetic
gauge invariance the photonic couplings $f_1^{\gamma}$ and 
$f_2^{\gamma}$ are related by 
$f_1^{\gamma}= \eta f_2^{\gamma}$, thus reducing the total number of unknowns to
thirteen . 


\setcounter{equation}{0}
\section{Polarized cross-sections}

In this section we will
augment the previous set of observables, which were 
projected out of the unpolarized differential cross-section, 
with three additional observables obtained from 
measurements of polarized total cross-sections. 
As a first step we will only compute in this section 
the polarized cross-sections obtained 
using non standard couplings that 
separately respect $C$ and $P$, i.e., we only retain the first 
three  $f_1^V,f_2^V,f_3^V$. 
In order to calculate the polarized cross-sections we 
define the following basic matrix elements for the production of two $W$s 
with definite helicity from polarized $e^-e^+$ beams. For massless 
electrons 
the helicity of the positron is opposite to the polarization 
of the electron: $\sigma_1=-\sigma_2=\sigma$.  
Three basic matrix elements are defined,  one for each  of the 
first three terms of the trilinear gauge vertex
in Eq.(\ref{anomalia}), and a fourth one 
for the neutrino exchange $t$-channel graph (Fig.1c):
\bea
{\cal M}_1^{\sigma}(\lambda_1,\lambda_2) &=& 
\left[\bar{v}(k_2,-\sigma)\slp_2 P_{\sigma}
u(k_1,\sigma) \right] 
\left( \epsilon_{\lambda_1}(p_1)\cdot
\epsilon_{\lambda_2}(p_2)\right) \nonumber\\
{\cal M}_2^{\sigma}(\lambda_1,\lambda_2)&=&
\left[\bar{v}(k_2,-\sigma)\slp_1
P_{\sigma} u(k_1,\sigma) \right] 
\left(p_1\cdot\epsilon_{\lambda_2}(p_2)\right)
\left(p_2\cdot\epsilon_{\lambda_1}(p_1)\right)
\frac{1}{2M_W^2} \nonumber\\
{\cal M}_3^{\sigma}(\lambda_1,\lambda_2)&=& 
\bar{v}(k_2,-\sigma)\left[
\esl_{\lambda_1}(p_1)\left(p_1\cdot\epsilon_{\lambda_2}(p_2)
\right)-\esl_{\lambda_2}(p_2)
 \left(p_2\cdot\epsilon_{\lambda_1}(p_1)\right)
\right]P_{\sigma} u(k_1,\sigma)\nonumber\\
{\cal M}_4^{\sigma}(\lambda_1,\lambda_2)&=& 
\bar{v}(k_2,-\sigma)\esl_{\lambda_2}(p_2)
 (\ksl_1-\slp_1)\esl_{\lambda_1}(p_1)
 P_{\sigma}u(k_1,\sigma)
\eea 
where the helicity projectors are given by 
\be
P_{\pm} = \frac{1\pm\gamma_5}{2}~.
\ee
We now establish contact with the notation of the 
previous section and that of \cite{PP}.  
In terms of the  basic matrix elements, defined above,  
the amplitudes corresponding to the three graphs 
of the $W$ pair-production process in Fig. 1 
are expressed as :
\bea
{\cal M}_{\gamma}^{\sigma}(\lambda_1,\lambda_2)&=& 
\frac{2s_w^2}{s}\left[(1+f_1^{\gamma})
{\cal M}_1^{\sigma}(\lambda_1,\lambda_2)
+f_2^{\gamma}{\cal M}_2^{\sigma}(\lambda_1,\lambda_2)
+(1+f_3^{\gamma}){\cal M}_3^{\sigma}(\lambda_1,\lambda_2)
\right] \nonumber \\
{\cal M}_{Z}^{\sigma}(\lambda_1,\lambda_2)&=& 
\frac{2g_{\sigma}}{s-M_Z^2}\left[(1+f_1^{Z})
{\cal M}_1^{\sigma}(\lambda_1,\lambda_2)
+f_2^{Z}{\cal M}_2^{\sigma}(\lambda_1,\lambda_2)
+(1+f_3^{Z}){\cal M}_3^{\sigma}(\lambda_1,\lambda_2)
\right] \nonumber \\
{\cal M}_{\nu}^{\sigma}(\lambda_1,\lambda_2)&=&
 -\frac{1}{2t}
{\cal M}_4^{\sigma}(\lambda_1,\lambda_2)\delta_{\sigma -}
\eea
where an overall coupling constant factor of $ig^2$  
has been pulled out,  
the left and right handed couplings of the electron 
with the $Z$ boson are given by  
\be 
g_+= v-a~,~~~ g_-=v+a,
\ee
and the Kronecker $\delta$ 
($\delta_{--} = \delta_{++} = 1$, $\delta_{-+}= \delta_{+-} = 0$)
in the neutrino graph appears due to 
the fact that the $W$ bosons couple only to left 
handed electrons.

The full amplitude can then be cast in the form:
\bea
{\cal M}^{\sigma}(\lambda_1,\lambda_2)&=&  
{\cal M}_{\gamma}^{\sigma}(\lambda_1,\lambda_2)
+{\cal M}_{Z}^{\sigma}(\lambda_1,\lambda_2)
+{\cal M}_{\nu}^{\sigma}(\lambda_1,\lambda_2)
\nonumber\\
&=&\frac{1}{s}\sum_{i=1}^{4}
F^{\sigma}_i{\cal M}_i^{\sigma}(\lambda_1,\lambda_2)
\eea
where 
\bea
F_i^{\sigma}&=& 
2s_w^2(1+f_i^{\gamma})+2g_{\sigma}u(1+f_i^{Z})
,~~~~~~~~~~~~ \mbox{for} ~i=1,3 \nonumber\\
F_2^{\sigma}&=& 2s_w^2Q_ef_2^{\gamma}
+2g_{\sigma}uf_2^{Z}  ~,\nonumber\\
F_4^{\sigma}&=&-\frac{s}{2t}\delta_{\sigma -} ~,
\eea
are explicit functions of the anomalous couplings.  

We then calculate the total cross-sections for 
the production of: (i) two transversely polarized $W$s 
denoted by $\sigma_T$, (ii) of two longitudinally polarized $W$s 
called $\sigma_L$ and (iii)  one transverse and one longitudinal $W$, 
denoted $\sigma_M$.  
We will present their explicit expressions 
in terms of arbitrary trilinear gauge couplings 
$f^V_1,f^V_2,f_3^V$. 
The relevant differential polarized 
cross-sections are defined by:
\bea
\frac{d\sigma_{T}}{dx} 
&=& \frac{1}{2s}\frac{\beta}{16\pi}\frac{g^4}{4} 
\sum_{\sigma,\lambda_1,\lambda_2=\pm}
|{\cal M}(\sigma,\lambda_1,\lambda_2)|^2 ~,
 \nonumber\\
\frac{d\sigma_{M}}{dx} 
&=& \frac{1}{2s}\frac{\beta}{16\pi}\frac{g^4}{4} 
\sum_{\sigma,\lambda=\pm}\left[ 
|{\cal M}(\sigma,\lambda,0)|^2 
+|{\cal M}(\sigma,0,\lambda)|^2 \right] ~,
\nonumber\\  
\frac{d\sigma_{L}}{dx} 
&=& \frac{1}{2s}\frac{\beta}{16\pi}\frac{g^4}{4} 
\sum_{\sigma=\pm}|{\cal M}(\sigma,0,0)|^2  ~.
\eea
These are calculated in a straightforward manner using 
the expressions of the basic matrix elements
for the different polarization combinations. 
The non-vanishing amplitudes are
explicitly given below:

\underline{TT} 
\bea
{\cal M}_1(\sigma,\pm,\pm)&=&
 -\frac{\beta s}{2}\sqrt{1-x^2} ~, \nonumber\\
{\cal M}_4(\sigma,\pm,\pm)&=&
-\frac{\beta s}{2}(x-\beta)\sqrt{1-x^2}  ~,\nonumber\\
{\cal M}_4(\sigma,\pm,\mp)&=&
-\frac{s}{2}(x\mp\sigma)\sqrt{1-x^2} ~, 
\eea

\underline{TL} 
\bea
{\cal M}_3(\sigma,\pm,0)={\cal M}^{\sigma}_3(0,\mp)&=& 
\sqrt{2\eta}\frac{\beta s}{2}(x\mp\sigma) ~,\nonumber\\ 
{\cal M}_4(\sigma,\pm,0)={\cal M}^{\sigma}_4(0,\mp)&=&
-\sqrt{2\eta}\frac{s}{4}
[2(\beta-x)\mp\sigma/\eta)(x\mp\sigma) ~,
\eea

\underline{LL} 
\bea
{\cal M}_1(\sigma,0,0)&=& 
-\frac{\beta s(2\eta-1)}{2}\sqrt{1-x^2} ~,\nonumber \\
{\cal M}_2(\sigma,0,0)&=& 
\beta \eta s(\eta-1)\sqrt{1-x^2}  ~,\nonumber\\
{\cal M}_3(\sigma,0,0)&=&
 2\beta \eta s\sqrt{1-x^2}  ~,\nonumber\\
{\cal M}_4(\sigma,0,0)&=& 
-\frac{s}{2}[\beta(2\eta+1)-2\eta x]\sqrt{1-x^2}  ~.
\eea
Notice that, as is well known \cite{Bil}, the transverse 
cross-section $\sigma_T$ receives anomalous contributions 
only from  $f_1^V$, 
whilst  
$\sigma_M$  only from $f_3^V$. 
Finally, the longitudinal cross-section  $\sigma_L$ 
depends on all six anomalous 
form-factors $f_1^V,~f_2^V,~f_3^V$. 

Using the expressions given above, 
we first compute the differential 
cross-sections and, as a check, we verify that by combining 
all three we obtain again the results of the previous 
section. After performing the angular integration in 
order to obtain the total cross-sections we 
also check, by setting $f_i^V \to 0$, that our SM result 
agrees with the polarized cross-sections 
presented in \cite{BD}. 
After these basic checks of our calculation, we subtract 
 the SM contribution to obtain
three new observables 
\bea
\sigma_{5} &\equiv & \left[\frac{128\pi s}{g^4\beta} \right]
(\sigma^{exp}_{TT}-\sigma_{TT}^0)  ~,\nonumber\\
\sigma_{6} &\equiv & \left[\frac{128\pi s}{g^4\beta} \right]
(\sigma^{exp}_{LT}-\sigma_{LT}^0)  ~,\nonumber\\
\sigma_{7} &\equiv &\left[\frac{128\pi s}{g^4\beta} \right]
( \sigma^{exp}_{LL}-\sigma_{LL}^0) ~.
\eea
Setting for convenience  
\be
{\cal L}\equiv\ln\left(\frac{1+\beta}{1-\beta}\right)  ~,
\ee
\bea
\tau_1 & \equiv &
-\frac{\eta}{\beta^2}+1+\frac{8\eta}{3}+\beta^2(1+\eta) ~,
\nonumber\\
\tau_2 & \equiv &
\eta-\beta^2\left(1+\frac{8\eta}{3}\right)-3\beta^4(1+\eta) ~,\nonumber\\
\tau_3 & \equiv & \frac{1}{\beta^2}-\frac{8}{3}-\beta^2 ~,\nonumber\\
\tau_4 & \equiv &
\frac{4}{\beta^2 }+\frac{16}{3}+12\beta^2  ~,
\eea
\bea
\begin{array}{ll}
Q^5 \equiv \frac{16\beta^2}{3}~, &
Q^6 \equiv \frac{128\eta\beta^2}{3}~, \\
Q^7_A \equiv  \frac{8}{3}\beta^2(2\eta-1)^2 ~, &
Q^7_B \equiv \frac{128}{3}\beta^6\eta^4~,\\
Q^7_C \equiv \frac{128}{3}\beta^2\eta^2~, &
Q^7_D \equiv \frac{64}{3}\beta^4(2\eta-1)\eta^2~,\\
Q^7_E \equiv \frac{64}{3}\beta^2(2\eta-1)\eta~,&
Q^7_F \equiv \frac{256}{3}\beta^4\eta^3~,
\end{array} 
\eea
the polarized observables $\sigma_5,\sigma_6$, and $\sigma_7$
are given by 
\bea
\sigma_{5}&=& L^5_1\ f_1^{\gamma} +   L^5_8\ f_1^{Z}
+Q^5_{[1][1]}\ \left(f_1^{\gamma}\right)^2 
+Q^5_{[1][8]} \ f_1^{\gamma}f_3^{Z}
 +Q^5_{[8][8]} \ \left(f_1^{Z}\right)^2  ~,
\nonumber\\
\sigma_{6}&=& L^6_3\ f_3^{\gamma} +   L^6_{10}\ f_3^{Z}
+Q^6_{[3][3]}\ \left(f_3^{\gamma}\right)^2 
+Q^6_{[3][10]} \ f_3^{\gamma}f_3^{Z}
 +Q^6_{[10][10]} \ \left(f_3^{Z}\right)^2  ~,
\nonumber\\
\sigma_{7}&=& \sum_k L^7_k \ c_k + \sum_k \sum_{l\geq k}Q^7_{[k][l]} c_kc_l
\label{sig7} ~,
\eea
and the various coefficients appearing in Eq.(\ref{sig7}) are 
explicitly given below:

\bea
L^5_1&=&s_w^2\left[\tau_3+
\frac{32\beta^2}{3} (s_w^2 +vu)
-\frac{1}{2\beta^3\eta^3}{\cal L}\right] ~, 
\nonumber\\ 
L^5_8&=&u\left[ (v+a)\tau_3+
\frac{32\beta^2}{3} \left(s_w^2 v +ru\right)
-\frac{(v+a)}{2\beta^3\eta^3}{\cal L}
\right] ~,\nonumber\\
Q^5_{[1][1]} &=& s_w^4 \ Q^5~, \nonumber\\
Q^5_{[1][8]} &=& 2s_w^2vu \ Q^5~, \nonumber\\
Q^5_{[8][8]} &=& ru^2 \ Q^5 ~,
\eea 

\bea
L^6_3 &=& s_w^2 \eta \left[
-\tau_4
+\frac{256\beta^2}{3} (s_w^2+vu )
+\frac{2(1+3\beta^2)}{\beta^3\eta^2}{\cal L}\right] f_3^{\gamma} ~,
 \nonumber\\
L^6_{10}&=&u\eta\left[ 
-(v+a) \tau_4
+\frac{256\beta^2}{3}(vs_w^2+ru )
+(v+a)\frac{2(1+3\beta^2)}{\beta^3\eta^2}{\cal L}
\right] ~, \nonumber\\
Q^6_{[3][3]} &=& s_w^4 \ Q^6~,~\nonumber\\ 
Q^6_{[3][10]} &=& 2s_w^2vu \ Q^6~,~ \nonumber\\
Q^6_{[10][10]} &=& ru^2 \ Q^6 ~,
\eea

\bea
L^7_1&=&          
s_w^2(2\eta-1)\left[
\tau_1
 -(s_w^2+vu)\frac{16}{3}\beta^2(2\eta+1)
+\frac{1}{2\eta^3\beta^3}{\cal L}
\right]  ~, \nonumber\\
L^7_8&=&u(2\eta-1)
\left[
(v+a)\tau_1
 -(s_w^2v+ru)\frac{16}{3}\beta^2(2\eta+1)
+\frac{(v+a)}{2\eta^3\beta^3}{\cal L}
\right]  ~,\nonumber\\
L^7_2&=&                              
s_w^2\left[4\eta^2\tau_2
+(s_w^2+vu)\frac{64}{3}\beta^4\eta(2\eta+1)
-\frac{2}{\eta\beta}{\cal L} \right]  ~,\nonumber\\
L^7_9&=&u\left[(v+a)4\eta^2
\tau_2
+(s_w^2v+ru)\frac{64}{3}\beta^4\eta(2\eta+1)
-\frac{2(v+a)}{\eta\beta}{\cal L} \right]  ~,\nonumber\\
L^7_3&=&                   
-4s_w^2 \left[
\tau_1
 -(s_w^2+vu)\frac{16}{3}\beta^2(2\eta+1)
+\frac{1}{2\eta^3\beta^3}{\cal L}
\right]  ~,\nonumber\\
L^7_{10}&=&-4u  \left[
(v+a)\tau_1
 -(s_w^2v+ru)\frac{16}{3}\beta^2(2\eta+1)
+\frac{(v+a)}{2\eta^3\beta^3}{\cal L}
\right]  ~,\nonumber\\
Q^7_{[1][1]} &=&s_w^4Q^7_A ~,\nonumber\\
Q^7_{[1][8]} &=& 2s_w^2vu Q^7_A ~,\nonumber\\
Q^7_{[8][8]} &=& ru^2 Q^7_A ~,\nonumber\\
Q^7_{[2][2]} &=& s_w^4 Q^7_B~,\nonumber\\
Q^7_{[2][9]} &=& 2s_w^2vu Q^7_B  ~,\nonumber\\
Q^7_{[2][9]} &=& ru^2 Q^7_B ~, \nonumber\\
Q^7_{[3][3]} &=& s_w^4 Q^7_C  ~,\nonumber\\
Q^7_{[3][10]} &=& 2s_w^2vu Q^7_C  ~,\nonumber\\
Q^7_{[10][10]} &=& ru^2 Q^7_C  ~,\nonumber\\
Q^7_{[1][2]} &=& -s_w^4 Q^7_D ~,\nonumber\\
Q^7_{[1][9]} &=& Q^7_{[2][8]}= s_w^2vu Q^7_D  ~,\nonumber\\
Q^7_{[8][9]} &=& ru^2 Q^7_D ~,\nonumber\\
Q^7_{[1][3]} &=& -s_w^4 Q^7_E  ~,\nonumber\\
Q^7_{[1][10]} &=& Q^7_{[3][8]} = s_w^2vu Q^7_E  ~,\nonumber\\
 Q^7_{[8][10]} &=& ru^2 Q^7_E  ~,\nonumber\\
Q^7_{[2][3]} &=&-s_w^4 Q^7_F  ~,\nonumber\\
Q^7_{[2][10]} &=& Q^7_{[3][9]}= s_w^2vu Q^7_F  ~,\nonumber\\
 Q^7_{[9][10]} &=& ru^2 Q^7_F~.
\eea

\section{$C$ and $P$ conserving couplings.}

In what follows we will focus on the special case where 
 all anomalous couplings satisfy
the individual discrete symmetries $C$, $P$, and $T$, 
i.e. we assume that $f_4^V=f_5^V=f_6^V=f_7^V=0$.
Then, the polarized $\sigma_i$ for $i=5,6,7$ are given in 
Eqs.(\ref{sig7}), while 
the corresponding unpolarized $\sigma_i$ for 
$i=1,...,4$ 
assume the following form:

\bea
\sigma_1 &=& L^{1}_{3}f_3^{\gamma}+L^1_{10}f_3^{Z}+ 
Q^1_{[3][3]}(f_3^{\gamma})^2
+ Q^1_{[3][10]}f_3^{\gamma}f_3^{Z}+Q^1_{[10][10]}(f_3^{Z})^2 ~,
\nonumber\\
\sigma_2 &=& L^2_{1}f_1^{\gamma}+L^2_{2}f_2^{\gamma}
+L^2_{3}f_3^{\gamma}+L^2_{8}f_1^{Z}
+L^2_{9}f_2^{Z}+L^2_{10}f_3^{Z}
            + Q^2_{[1][1]}(f_1^{\gamma})^2
            + Q^2_{[1][2]}f_1^{\gamma}f_2^{\gamma}  
            + Q^2_{[1][3]}f_1^{\gamma}f_3^{\gamma}\nonumber\\  
&&            + Q^2_{[1][8]}f_1^{\gamma}f_1^{Z} 
            + Q^2_{[1][9]}f_1^{\gamma}f_2^{Z}  
            + Q^2_{[1][10]}f_1^{\gamma}f_3^{Z} 
           + Q^2_{[2][2]}(f_2^{\gamma})^2 
            + Q^2_{[2][3]}f_2^{\gamma}f_3^{\gamma} 
            + Q^2_{[2][8]}f_2^{\gamma}f_1^{Z} \nonumber\\
&&            + Q^2_{[2][9]}f_2^{\gamma}f_2^{Z} 
            + Q^2_{[2][10]}f_2^{\gamma}f_3^{Z}  
            + Q^2_{[3][3]}(f_3^{\gamma})^2 
            + Q^2_{[3][8]}f_3^{\gamma}f_1^{Z} 
            + Q^2_{[3][9]}f_3^{\gamma}f_2^{Z} 
            + Q^2_{[3][10]}f_3^{\gamma}f_3^{Z}\nonumber\\ 
&&            + Q^2_{[8][8]}(f_1^{Z})^2  
            + Q^2_{[8][9]}f_1^{Z}f_2^{Z} 
            + Q^2_{[8][10]}f_1^{Z}f_3^{Z}
            + Q^2_{[9][9]}(f_2^{Z})^2 
            + Q^2_{[9][10]}f_2^{Z}f_3^{Z}
            + Q^2_{[10][10]}(f_3^{Z})^2 ~,
\nonumber\\
\sigma_3 &=& L^3_{1} f_1^{\gamma}  + L^3_{2} f_2^{\gamma}  + 
L^3_{8} f_1^{Z}  +L^3_{9} f_2^{Z} ~,
\nonumber\\
\sigma_4 &=& L^4_{3} f_3^{\gamma}  + L^4_{10} f_3^{Z} ~.
\label{CPT}
\eea 

The following comments are in order:

(i) 
Notice that the expressions for $\sigma_3$ and $\sigma_4$ 
receive no quadratic contributions and are therefore
identical to those presented in \cite{PP}. 

(ii) The expressions for $\sigma_1$ and $\sigma_4$ constitute a system 
of two equations with two unknowns, $f_3^{\gamma}$ and $f_3^{Z}$, 
as was the case in \cite{PP}, but now the unknown quantities 
appear quadratically in $\sigma_1$. As we will see in a moment, 
one of the results of
this is that the degeneracy between the two systems is
improved.

(iii)
By measuring the polarized quantities, 
one would arrive at a system of
seven equations for the six unknown form-factors. 
In fact, the system separates
into two sub-systems:  One sub-system of three equations 
$\{ \sigma_1, \sigma_4, \sigma_{6}\}$ with two unknowns 
$\{ f_3^{\gamma},f_3^{Z} \}$, and one sub-system of
the remaining four equations involving all six unknowns.
One could then attempt a global solution, or use the first sub-system
to determine $f_3^{\gamma}$ and $f_3^{Z}$, and use their values
as input in the other. Notice also that the fact that we have
three equations for $f_3^{\gamma}$ and $f_3^{Z}$ may reduce or 
eliminate completely the ambiguities in 
determining them which originate from the
quadratic nature of these equations \cite{RLS}.

Given that 
$\{ \sigma_1, \sigma_4, \sigma_{6}\}$ constitute an independent 
sub-system, it is interesting to carry out 
an elementary study of their correlations,
at least within the context
of a simple model simulating the statistical behaviour of the anomalous
couplings. 
Such a study is useful since 
the three observables involved appear to be 
intrinsically different in nature, at least in as far as their
inclusiveness is concerned: $ \sigma_1$ and $\sigma_4$ originate
from convolutions of the unpolarized differential cross-section with the
corresponding projective polynomials, whereas $\sigma_{6}$
originates from selecting those specific events 
of the full cross-section that 
correspond to longitudinally polarized  $W$ bosons.
We will assume that the two couplings  
$f_3^{\gamma}\equiv z_1$ and $f_3^{Z}\equiv z_2$
obey independently a normal (Gaussian) probability distribution,
with mean $\mu_i$ and variance $ \delta_i^2$, i.e. 
\be
p_{i}(z_i, \mu_i, \delta_i^2) = \frac{1}{\delta_i (2\pi)^{\frac{1}{2}}}
\exp \Bigg [ -\frac{(z_i- \mu_i)^2}{2\delta_i^2}\Bigg ]~.  
\ee
Then, the expectation value
$\langle \sigma_i\rangle$ of the observable $\sigma_i,~i=1,4,6 $ 
is given by
\be
\langle \sigma_i\rangle = \prod\limits_{j=1}^{2}
\int_{-\infty}^{+\infty} 
[dz_{j}] p_j \sigma_i~,
\ee
the corresponding covariance matrix by
\be
V_{ij}= \langle \sigma_i \sigma_j \rangle -
\langle \sigma_i \rangle \langle \sigma_j \rangle ~,
\ee
and the correlations $r_{ij}$  by
\be
r_{ij} = \frac{V_{ij}}{V_{ii}^{1/2}V_{jj}^{1/2}}~.
\label{rij}
\ee

We will next assume that the Gaussian distribution is
peaked around the SM 
values of the couplings, i.e. $\mu_i=0$, 
and will use the elementary results
\be
\int_{-\infty}^{+\infty}[dz_{i}]
 z_i p_{i}^{(0)} =0,~~
\int_{-\infty}^{+\infty}[dz_{i}]z_i^2
 p_{i}^{(0)} =\delta_i^2,~~
\int_{-\infty}^{+\infty}[dz_{i}] z_i^4
 p_{i}^{(0)} =\frac{3}{4}\delta_i^4 ~,
\ee
where $p_{i}^{(0)}\equiv p_{i}(z_i,0, \delta_i^2)$.

We next study the correlations $r_{ij}$
in the absence and presence of quadratic corrections.
We assume for simplicity that
$\delta_1=\delta_2 =\delta$; actually, the final results 
do not depend on $\delta$, which cancels out 
when forming the ratios in Eq.(\ref{rij}).
The results   
for some characteristic values
of the center-of-mass energy $\sqrt s$ are given in 
Table 1 and Table 2, respectively.

\begin{center}
\begin{tabular}{|c|l|l|l|l|l|}\hline
$\sqrt s$ (GeV) & 180 & 200& 250 & 300 & 500 \\ \hline\hline  
$r_{14}$ & -0.999  & -0.998 & -0.995 & -0.992 &-0.988 \\ \hline\hline
$r_{16}$ & 0.619& 0.763 &0.894 & 0.936 & 0.975\\ \hline\hline
$r_{46}$ & -0.588& -0.719 & -0.842 &-0.885 & -0.931 \\ \hline\hline
\end{tabular}
\end{center}
\noindent  {\bf Table 1} :
The correlation coefficients $\rho_{ij}$ as a function of  
$\sqrt s$ in the
absence of quadratic corrections.
\vspace*{0.5cm}
\medskip

\begin{center}
\begin{tabular}{|c|l|l|l|l|l|}\hline
$\sqrt s$ (GeV) & 180 & 200& 250 & 300 & 500 \\ \hline\hline  
$r_{14}$ &-0.992 & -0.970 & -0.850 & -0.686 & -0.267 \\ \hline\hline
$r_{16}$ &0.333 & 0.482 & 0.724 & 0.859 & 0.986 \\ \hline\hline
$r_{46}$ &-0.215 &-0.254 & -0.253 & -0.218 & -0.107 \\ \hline\hline
\end{tabular}
\end{center}
\noindent  {\bf Table 2
} :
The correlation coefficients $\rho_{ij}$ as a function of  
$\sqrt s$ in the
presence of quadratic corrections.
\vspace*{0.5cm}
\medskip

We notice that for all values of $\sqrt s$ the inclusion of
the quadratic terms leads to lower values for the correlations.
The elementary analysis presented above can be easily generalized
to include all seven observables, 
thus constructing the full correlation matrix.

(iv)
We note that $f_1$ and $f_3$ are 
enhanced at the production threshold ($\beta \to 0$) due to the 
factors $1/\beta^2$ which survive in their coefficients 
($\tau_1,~\tau_3,~\tau_4$) 
in the polarized observables (remember that there is an  
overall prefactor $\propto \beta$ stemming  from phase space).
This enhancement cancels
 in the total cross-section $\sigma=\sigma_T+\sigma_M+\sigma_L$; this 
known fact furnishes an additional useful check of our calculation. 
Evidently, 
the measurement of the polarized cross-sections will be 
more sensitive to  the $f_1$ and $f_3$ form-factors 
at the low-energy LEP2 runs.

(v) Finally one of the unknown photonic deviations $f_2^{\gamma}$ can be 
completely eliminated by resorting to electromagnetic $U(1)$ gauge-invariance;   
the latter imposes  
on the deviation form-factors $f_1^{\gamma}$ and $f_2^{\gamma}$
the relation
\be
f_1^{\gamma}=\eta f_2^{\gamma}. 
\label{U1f1f2}
\ee
Thus, the number of unknown form-factors appearing in 
Eq.(\ref{CPT}) and  Eq.(\ref{sig7})
will be reduced down to five, 
a fact which should
restrict even further
any ambiguities stemming from the quadratic nature of the equations. 

\section{Conclusions}

We have obtained explicit expressions of the unpolarized differential 
cross-section for the production of an on shell $W$ pair keeping the most
general structure for the triple gauge boson vertices, namely 
all fourteen different 
form-factors which parametrize the deviations  
from the tree-level SM trilinear gauge vertex. 
The above explicit result, which contains all linear and quadratic terms 
in the anomalous couplings, demonstrates that the unpolarized differential 
cross-section can be expressed in terms of four polynomials in the cosine 
of the  center-of-mass scattering angle, of maximum degree 3, 
linearly independent and identical to those  obtained in the simpler 
case where only linear terms of the $C$,$P$ and $T$ conserving couplings were 
kept. The corresponding coefficients multiplying
these polynomials can be projected 
out from the differential cross-section;
they constitute a set of four
observables, whose measurement imposes experimental constraints 
on the anomalous couplings. 

Furthermore, we have augmented the 
aforementioned set of observables by three additional ones,
which correspond to the total cross-sections 
 for obtaining in the final state
$W$ bosons with fixed polarization 
(both transverse, both longitudinal, one transverse and one longitudinal)
in the presence of $C$,$P$ and $T$ conserving anomalous couplings.
The experimental value of these observables can be extracted
from measurements of the polarization of the final state $W$ bosons.

The proposed observables comprise 
a set of seven quadratic equations containing 
fourteen unknowns, which could be simultaneously fitted in order to put 
global constraints on all anomalous couplings.  
Alternatively, one could focus exclusively on the
subset of anomalous couplings which separately respect 
the $C$, $P$, and $T$ discrete symmetries, thus arriving at an
over-constrained system; the latter could be used in order to
eliminate possible algebraic ambiguities in the determination of the
above couplings, or reduce their correlations.
Imposing in addition 
electromagnetic gauge invariance one can further
restrict the number of unknowns in the above system.
 
It would be interesting to see how this method  
responds first to simulated and subsequently to real data.
A first step towards a full realization of the method has been 
recently reported, focussing mainly on aspects related to its 
experimental feasibility \cite{ww99}.

\vspace{0.7cm}\noindent {\bf  Acknowledgments.}  
The authors  
thank R.L. Sekulin for suggesting to them
the inclusion of the polarized cross-sections 
and for various useful comments, 
and E.Sanchez for numerous informative discussions.

\newpage
\centerline{FIGURE}

\bigskip

\begin{center}
\begin{picture}(400,130)(0,20)

\put( 34, 58){\makebox(0,0)[c]{\small $e^{+}(k_{2},-\sigma)$}}
\put( 34,144){\makebox(0,0)[c]{\small $e^{-}(k_{1},\sigma)$}}
\ArrowLine( 40, 70)( 55,100)
\ArrowLine( 40,130)( 55,100)
\put( 73,115){\makebox(0,0)[c]{\small $\gamma$}}
\Photon( 55,100)( 85,100){4}{3}
\Photon( 85,100)(100, 70){4}{3}
\Photon( 85,100)(100,130){4}{3}
\put(111, 58){\makebox(0,0)[c]{\small $W^{+}(p_{2},\lambda_{2})$}}
\put(111,144){\makebox(0,0)[c]{\small $W^{-}(p_{1},\lambda_{1})$}}
\put( 70,40){\makebox(0,0)[c]{(a)}}
\GCirc(85,100){9}{0.8}
\Text(85,100)[]{$\Gamma^{\gamma}$}
 
\ArrowLine(170, 70)(185,100)
\ArrowLine(170,130)(185,100)
\put(203,115){\makebox(0,0)[c]{\small $Z$}}
\Photon(185,100)(215,100){4}{3}
\Photon(215,100)(230, 70){4}{3}
\Photon(215,100)(230,130){4}{3}
\put(200,40){\makebox(0,0)[c]{(b)}}
\GCirc(215,100){9}{0.8}
\Text(215,100)[]{$\Gamma^{Z}$}

\ArrowLine(300, 70)(330, 85)
\ArrowLine(300,130)(330,115)
\put(322,100){\makebox(0,0)[c]{\small $\nu_{e}$}}
\Line(330, 85)(330,115)
\Photon(330, 85)(360, 70){-4}{3}
\ArrowLine(345, 77.5)(347, 79)
\Photon(330,115)(360,130){4}{3}
\ArrowLine(345,122.5)(347,121)
\put(330,40){\makebox(0,0)[c]{(c)}}

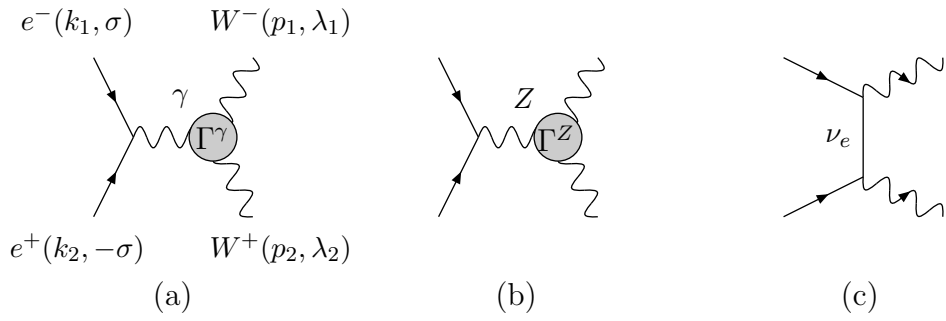
\Text(200,0)[]{Fig.~1: The process
$e^{+}e^{-} \rightarrow W^{+}W^{-}$ at tree level, including
anomalous gauge boson couplings.}

\end{picture}
\end{center}

\end{document}